\documentstyle[12pt,a4]{article}
\author{L.A. Leonovich, A.T. Altynsev, V.V. Grechnev, E. L. Afraimovich \\
        Institute of Solar-Terrestrial Physics SD RAS,\\
        p.~o.~box~4026, Irkutsk, 664033, Russia\\
        fax: +7 3952 462557; e-mail:~lal@iszf.irk.ru}

\title{Ionospheric effects of the solar flares as deduced from
       global GPS network data}
\date{}
\begin{document}
\sloppy
\maketitle
\begin{abstract}

Results derived from analysing the ionosphere response to faint
and bright solar flares are presented. The analysis used
technology of a global detection of ionospheric effects from
solar flares as developed by the authors, on the basis of phase
measurements of the total electron content (TEC) in the
ionosphere using an international GPS network. The essence of the
method is that use is made of appropriate filtering and a
coherent processing of variations in the TEC which is determined
from GPS data, simultaneously for the entire set of visible GPS
satellites at all stations used in the analysis. This technique
is useful for identifying the ionospheric response to faint solar
flares (of X-ray class C) when the variation amplitude of the TEC
response to separate line-on-sight to GPS satellite is comparable
to the level of background fluctuations. The dependence of the
TEC variation response amplitude on the flares location on the
Sun is investigated.
\end{abstract}

\section{Introduction}

The enhancement of X-ray and ultraviolet (UV) emission that is
observed during chromospheric flares on the Sun immediately
causes an increase in electron density in the ionosphere. These
density variations are different for different altitudes and are
called Sudden Ionospheric Disturbances, SID, (Davies, 1990),
(Donnelly,1969). SIDs are generally recorded as the short wave
fadeout, SWF,( Stonehocker, 1970), sudden phase anomaly, SPA,
(Ohshio, 1971), sudden frequency deviation, SFD, (Donnelly,
1971), (Liu et al., 1996), sudden cosmic noise absorption, SCNA,
(Deshpande and Mitra, 1972), sudden enhancement/decrease of
atmospherics, SES, (Sao et al., 1970). Much research is devoted
to SID studies, among them a number of thorough reviews (Mitra,
1974), (Davies, 1990).

SFD are caused by an almost time-coincident increase in $E$-and
$F$-region electron densities at over 100 km altitudes covering
an area with the size comparable to or exceding that of the
region monitored by the system of HF radio paths (Davies, 1990),
(Donnelly,1969), (Liu et al., 1996). A limitation of this method
is the uncertainty in the spatial and altitude localization of
the UV flux effect, the inadequate number of paths, and the need
to use special-purpose equipment.

The effect of solar flares on the ionospheric $F$-region is also
manifested as a Sudden Increase of Total Electron Content, SITEC,
which was measured previously using continuously operating VHF
radio beacons on geostationary satellites (Mendillo et al., 1974),
(Davies, 1980). A serious limitation of methods based on
analyzing VHF signals from geostationary satellites is their
small and ever increasing (with the time) number and the
nonuniform distribution in longitude. Hence it is impossible to
make measurements in some geophysically interesting regions of
the globe, especially in high latitudes.

A further, highly informative, technique is the method of
Incoherent Scatter - IS (Mendillo et al., 1974), (Thome et al.,
1971). However, the practical implementation of the IS method
requires very sophisticated, expensive equipment. An added
limitation is inadequate time resolution. Since the relaxation
time of electron density in the $E$ and $F1$ regions is also less
than 5-10 min, most IS measurements lack time resolution needed
for the study of ionospheric effects of flares.

Consequently, none of the above-mentioned existing methods can
serve as an effective basis for the radio detection system to
provide a continuous, global SID monitoring with adequate
space-time resolution. Furthermore, the creation of these
facilities requires developing special purpose equipment,
including powerful radio transmitters contaminating the radio
environment. It is also significant that when using the existing
methods, the inadequate spatial aperture gives no way of deducing
the possible spatial inhomogeneity of the X-ray and UV flux.

The advent and evolution of a Global Positioning System (GPS) and
also the creation on its basis of widely branched networks of GPS
stations (at least 800 sites at the February of 2001, the data
from which are placed on the Internet) opened up a new era in
remote ionospheric sensing. High-precision measurements of the
TEC along the line-of-sight (LOS) between the receiver on the
ground and transmitters on the GPS system satellites covering the
reception zone are made using two-frequency multichannel
receivers of the GPS system at almost any point of the globe and
at any time simultaneously at two coherently coupled frequencies
$f_{1}=1575.42$ MHz and $f_{2}=1227.60$ MHz.

The sensitivity of phase measurements in the GPS system is
sufficient for detecting irregularities with an amplitude of up
to $10^{3}-10^{4}$ of the diurnal TEC variation. This makes it
possible to formulate the problem of detecting ionospheric
disturbances from different sources of artificial and natural
origins. The TEC unit (TECU) which is equal to $10^{16}$m$^{-2}$
and is commonly accepted in the literature, will be used
throughout the text.

Afraimovich et al. (2000a, 2000b, 2001) developed a novel
technology of a global detection of ionospheric effects from
solar flares and presented data from first GPS measurements of
global response of the ionosphere to powerful impulsive flares of
July 29, 1999, and December 28, 1999, were chosen to illustrate
the practical implementation of the proposed method. Authors
found that fluctuations of TEC, obtained by removing the linear
trend of TEC with a time window of about 5 min, are coherent for
all stations and LOS on the dayside of the Earth. The time
profile of TEC responses is similar to the time behavior of hard
X-ray emission variations during flares in the energy range 25-35
keV if the relaxation time of electron density disturbances in
the ionosphere of order 50-100 s is introduced. No such effect on
the nightside of the Earth has been detected yet.

The objective of this paper is to use this technology for
analysing the ionosphere response to faint and bright solar
flares.

\section{Processing of the data from the GPS network}
\label{TSE-sect-2}

Following is a brief outline of the global monitoring (detection)
technique for solar flares.  A physical groundwork for the method
is formed by the effect of fast change in electron density in the
Earth's ionosphere at the time of a flare simultaneously on the
entire sunlit surface. Essentially, the method implies using
appropriate filtering and a coherent processing of TEC variations
in the ionosphere simultaneously for the entire set of visible
(during a given time interval) GPS satellites (as many as 5-10
satellites) at all global GPS network stations used in the
analysis.

In detecting solar flares, the ionospheric response is virtually
simultaneous for all stations on the dayside of the globe within
the time resolution range of the GPS receivers (from 30 s to 0.1
s). Therefore, a coherent processing of TEC variations implies in
this case a simple addition of single TEC variations. The
detection sensitivity is determined by the ability to detect
typical signals of the ionospheric response to a solar flare
(leading edge duration, period, form, length) at the level of TEC
background fluctuations. Ionospheric irregularities are
characterized by a power spectrum, so that background
fluctuations will always be distinguished in the frequency range
of interest. However, background fluctuations are not correlated
in the case of beams to the satellite spaced by an amount
exceeding the typical irregularity size. With a typical length of
X-ray bursts and UV emission of solar flares of about 5-10 min,
the corresponding ionization irregularity size does normally not
exceed 30-50 km; hence the condition of a statistical
independence of TEC fluctuations at spaced beams is almost always
satisfied. Therefore, coherent summation of responses to a flare
on a set of LOS spaced throughout the dayside of the globe
permits the solar flare effect to be detected even when the
response amplitude on partial LOS is markedly smaller than the
noise level (background fluctuations).

The proposed procedure of coherent accumulation is essentially
equivalent to the operation of coincidence schemes which are
extensively used in X-ray and gamma-ray telescopes. If the SID
response and background fluctuations, respectively, are
considered to be the signal and noise, then as a consequence of a
statistical independence of background fluctuations the
signal/noise ratio when detecting the flare effect is increased
through a coherent processing by at least a factor of $\sqrt N$,
where $N$ is the number of LOS.

The GPS technology provides the means of estimating TEC
variations on the basis of phase measurements of TEC $I$ in each
of the spaced two-frequency GPS receivers using the formula
(Hofmann-Wellenhof et al., 1992), (Calais and Minster, 1996):

\begin{equation}
\label{SPE-eq-01}
I=\frac{1}{40{.}308}\frac{f^2_1f^2_2}{f^2_1-f^2_2}
                           [(L_1\lambda_1-L_2\lambda_2)+const+nL]
\end{equation}

where $L_1\lambda_1$ and $L_2\lambda_2$  are the increments of
the radio signal phase path caused by the phase delay in the
ionosphere (m); $L_1$ and $L_2$ stand for the number of complete
phase rotations, and $\lambda_1$ and $\lambda_2$ are the
wavelengths (m) for the frequencies $f_{1}$ and $f_{2}$,
respectively; $const$ is some unknown initial phase path (m); and
$nL$ is the error in determining the phase path (m).

Phase measurements in the GPS system are made with a high degree
of accuracy where the error in TEC determination for 30-second
averaging intervals does not exceed $10^{14}$m$^{-2}$, although
the initial value of TEC does remain unknown (Hofmann-Wellenhof et
al., 1992).

This permits ionization irregularities and wave processes in the
ionosphere to be detected over a wide range of amplitudes (as
large as $10^{-4}$ of the diurnal variation of TEC) and periods
(from several days to 5 min). The TEC unit, $TECU$, which is
equal to $10^{16}$ m${}^{-2}$ and is commonly accepted in the
literature, will be used throughout the text.

The data analysis was based on using the stations, for which the
loval time during the flare was within 10 to 17 LT. From 50 to
150 LOS were processed for each flare.

Primary data include series of slant values of TEC $I(t)$, as
well as the corresponding series of elevations $\theta(t)$ and
azimuths $\alpha(t)$ along LOS to the satellite calculated using
our developed CONVTEC program which converts the GPS system
standard RINEX-files on the INTERNET (Gurtner, 1993). The
determination of SID characteristics involves selecting
continuous series of $I(t)$ measurements of at least a one-hour
interval in length, which includes the time of the flare. Series
of elevations $\theta(t)$ and azimuths $\alpha(t)$ of the LOS are
used to determine the coordinates of subionospheric points. In
the case under consideration, all results were obtained for
elevations $\theta(t)$ larger than $30^\circ$.

The method of coherent summation of time derivatives of the
series of variations of the "vertical" TEC value was employed in
studying the ionospheric response to solar flares. Our choice of
the time derivative of TEC was motivated by the fact this
derivative permits us to get rid of a constant component in TEC
variations; furthermore, it reflects electron density variations
that are proportional to the flux of ionizing radiation.

The coherent summation of time derivatives of the series of
variations of the "vertical" TEC value  was made by the formula:

\begin{equation}
\label{SPE-eq-02} Sw= \sum^n_{i=1} dI(t)/dt_{i}  \times K_{i}
\end{equation}

where  $n$ is the number of LOS. The correction coefficient
$K_{i}$ is required for converting the slant TEC to an equivalent
"vertical" value (Klobuchar, 1986)

\begin{equation}
\label{GLOB-eq-03} K_i =  {\rm  cos}  \left[{\rm
arcsin}\left(\frac{R_z}{B_z  + h_{{\rm max}}}{\rm cos}
\hspace{0.1cm}  \theta_i\right) \right],
\end{equation}

where $R_{z}$ is Earth's radius; and $h_{max}$ is the height of
the ionospheric $F2$-layer maximum.

Next the trend determined as a polynomial on a corresponding time
interval is removed from the result (normalized to the number of
LOS) of the coherent summation of the time derivatives. After
that, the calculated time dependence $(Sd_{ex}(t))$ is integrated
in order to obtain the mean integral TEC increment $\triangle$
$I(t)$ on the time interval specified.

\begin{equation}
\label{SPE-eq-04} \triangle I(t) =
\int\limits_{t1}^{t2}Sd_{ex}(t) dt
\end{equation}

This technique is useful for identifying the ionospheric response
to faint solar flares (of X-ray class C) when the variation
amplitude of the TEC response to separate LOS is comparable to the
level of background fluctuations.

\section{Ionospheric response to faint solar flares }
\label{TSE-sect-3}

An example of a processing of the data for a faint solar flare
July 29, 1999 (C2.7/ SF, 11:11 GT, S16W11) is given in Figure 1.
One hundred LOS were processed for the analysis of this event.
Panels (a) and (b) present the typical time dependencies of TEC
variations for separate LOS, and their time derivatives. The BRUS
(PRN14, thick line) and BAHR (PRN29, thin line) stations are
taken as our example. It is apparent from these dependencies that
no response to the flare is distinguished in the TEC variations
and in their time derivatives for the individual LOS, because the
amplitude of the TEC response for the individual LOS is
comparable to the level of background fluctuations.

A response to the solar flare is clearly seen in the time
dependence (Figure 1c) which is a normalized result of a coherent
summation $Sd$ of the time derivatives of the TEC variations for
all LOS. Upon subtracting the trend determined as a polynomial of
degree 3 on the time interval 10:07-10:39 UT, the same curve (c)
is presented in Figure 1d as $Sd_{ex}(t)$. Next the calculated
time dependence was integrated over the time interval 10:07-10:39
UT to give the mean integral increment of TEC (Figure 1e, thick
line). A comparison of the resulting dependence with the values
of the soft X-ray emission flux (GOES-10) in the range 1-8 $\AA$
(Figure 1e, thin line) reveals that it has a more flattened form,
both in it rise and fall. A maximum in X-rays is about 6 minutes
ahead of that in TEC.

Examples of the application of our technology for the analysis of
the  ionospheric response to faint solar flares are given in
Figures 2 and 3.  Figure 2 gives the data processing results on
TEC  variations for solar flares :  July 29,  1999 (C2.2, 11{:}00
UT) on panels a, b, c, d,  and July 29,  1999 (C6.2/SN, 15{:}14
UT, N25E33 ) on panels e, f, g, h.  Figure  3 shows the results of
a data processing of TEC variations for solar flares of November
17,  1999 (C7.0/1N, 09{:}38 UT, S15W53) on panels a, b, c, d, and
November 11, 1999 (C5.0, 15{:}40 UT) on panels e, f, g, h.

\section{Ionospheric response to  bright flares }
\label{TSE-sect-4}

An example of a processing of the data for the bright solar flare
of July 14, 1998 (M4.6/1B, 12:59 UT, S23E20) is given in Figure 4.

Fifty LOS were used in the analysis of this event. Figure 4a
presents the time dependencies of hard X-ray emission (CGRO/Batse,
25-50 keV, thick line on panels a ) and of the UV line
(SOHO/SUMMER 171 $\AA$ , thin line) in arbitrary units (Aschwanden
et al., 1999). It should be noted that the time dependence of the
UV 171 $\AA$ line is more flattened, both in the rise and in the
fall, when compared with the hard X-ray emission characteristic.
The increase in the UV 171 $\AA$ line starts by about 1.8 minute
earlier, and the duration of its disturbance exceeds considerably
that of the hard X-ray emission disturbance.

Panel (b) presents the typical time dependencies of the TEC
variations for separate LOS. The AOML (PRN24, thick line) and ACSI
(PRN18, thin line) stations are taken as examples. A response to
the bright flare is clearly distinguished for separate LOS. The
normalized sum $Sd$ of the time derivatives of the TEC variations
for all LOS is presented in Figure 4c; panel (d) plots the same
curve (c), upon subtracting the trend determined as a polynomial
of degree 3 on the time interval 12:48-13:12 UT. Next the
resulting time dependence was integrated in order to obtain the
mean integral increment of TEC (Figure 4e). It might be well to
point out that the time dependence of the mean integral increment
of TEC has a more flattened form in the rise than the emission
flux characteristics; however, the onset time of its increase
coincides with that of hard X-ray emission, and is delayed by
about 1.8 minutes with respect to the UV 171 $\AA$ line.

A total of 11 events was processed. The class of X-ray flares was
from M4.5 to M7.4. It was found that the mean TEC variation
response in the ionosphere depends on the flare lecation on the
Sun (central meridian distance,  CMD) - Figure 5a.

\section{Discussion}
\label{GLOB-method}

Our results is consistent with the findings reported by (Donnelly,
1969), (Donnelly, 1971), (Donnelly, 1976), (Donnelly et al.,
1986a), where a study of extreme UV (EUV) flashes of solar flares
observed via SFD was made. In the cited references it was shown
that the relative strength of impulsive EUV emission from flares
decreases with increasing CMD and average peak frequency
deviation is also significantly lower for SFD's associated with
$H\alpha$ flares at large CMD. (Donnelly, 1971) is of the opinion
that percentage of $H\alpha$ flares with SFD's tends to decrease
for large CMD of $H\alpha$ flare location - Figure 5b. Similar
effects at the center and limb were observed in the ratio of EUV
flux to the concurrent hard X ray flux (Kane and Donnelly, 1971).
Using a fourth-order polynomial to fit the results in Figure 5b
with CMD in degrees (Donnelly, 1976) gives

\begin{equation} \label{GLOB-eq-05} A(CMD) \simeq 1 - 0.484 \cdot CMD
+ 0.001426 \cdot CMD^2-1.79 \cdot 10^{-5} CMD^3 + 7.43  \cdot
10^{-8} CMD^4 \end{equation}

Equation (5) implies that on the average the impulsive EUV
emission is more than an order of magnitude weaker for flares
near the solar limb than for flares at the central meridian.
(Donnelly, 1976) have assumed that it is result from the
low-lying nature of the $10^{4}K-10^{6}K$ flare source region and
from absorption of EUV emission in the surrounding cool nonflaring
atmosphere. Figure 5c presents the result of a modeling of the SFD
occurrence   probability at the time of the solar flare as a
function of CMD (solid line) in arbitrary units, as well as the
values amplitude of the TEC response in the ionosphere to solar
flares (in the range of X-ray class M4.9-M5.7 (dots), and
M6.6-M7.4 (grosses) as a function of CMD. The modeling used
equation (5). It figure suggests that the results of our
measurements do not contradict the conclusions drawn by Donnelly
(1976) that the relative strength of impulsive EUV emission from
flares decreases with increasing CMD. It should be noted that in
the case of solar flares whose class is similar in X-ray
emission, the dependence under study resembles cos(CMD) rather
than a polynomial of degree 4. The fitting cos(CMD) curve for
solar X-ray M4.4-M5.7 flares is plotted in Figure 5a (solid line)
and 5c (dashed line). This conclusion is consistent with the
findings reported by (Donnelly and Puga, 1990). In the cited
reference it was found the empirical curves of the average
dependence of active region emission on its CMD (Figure 6a) for
several wavelengs. Assume a quiet Sun plus one average active
region that starts at the center of the backside of the Sun and
rotates across the center of the solar disk with a 28-day period.

Fig. 6(b-e) presents our obtained dependencies of $\triangle I $
as a function of the angular distance of the flaring region from
the central solar meridian (CMD) for different classes of flares
in the X-ray range: X2.0-X5.7 (Fig. 6b, 6 flares), M2.0-M7.4 (Fig.
6c, 26 flares), M1.1 (Fig. 6d, 14 flares), and C2.7-C10 (Fig. 6e,
7 flares). Because of lack of space, we do not give here any
detailed characteristics of the flares. Our analysis was based on
using sets of GPS stations and LOS similar to those described in
the previous section. Fig. 6a presents the appropriate
dependencies of X-ray flux emission intensities (Mosher, 1979),
F10 cm (Riddle, 1969; Vauquois, 1955), and UV (Samain, 1979).

It is evident from Fig. 6 that the character of the response
amplitude corresponds to the behavior of ultraviolet lines. Hence
it follows that the main contribution to the ionospheric TEC
response to solar flares is made by the F region and the upper
part of the E region where solar ultraviolet radiation is
absorbed.

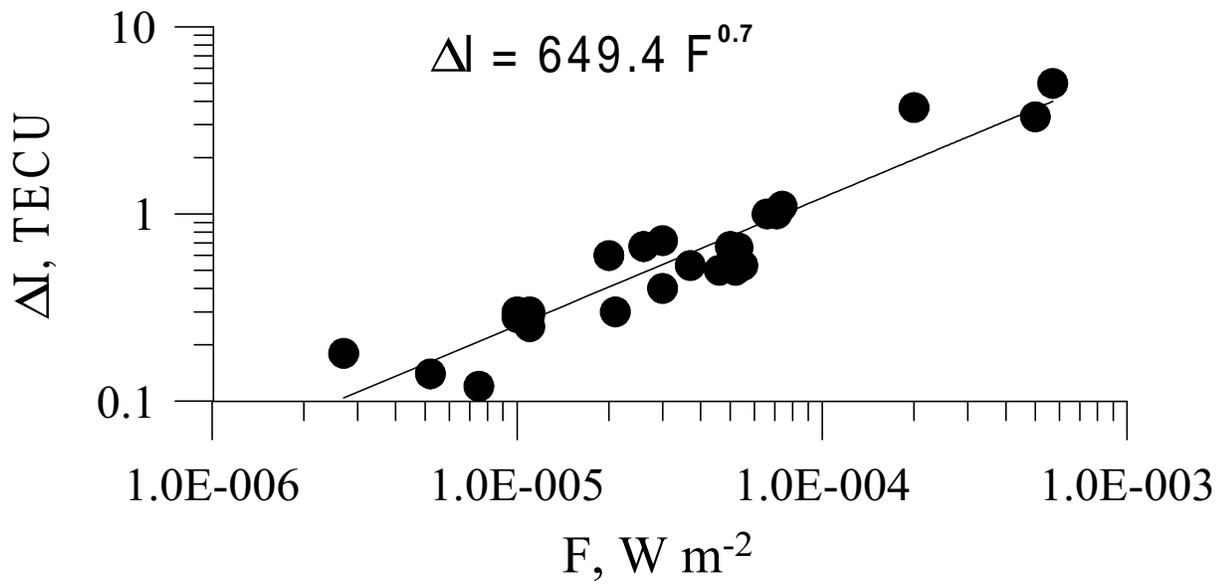
Fig. 7 plots, on a logarithmic scale, the dependence of the
response amplitude $\triangle I$ to solar flares as a function of
their peak power F in the X-ray range for flares located near the
center of the solar disk ($CMD<40^\circ$). This dependence
illustrates a wide dynamic range of the proposed method (three
orders in the flare power), and is quite well approximated by the
power-law function

\begin{equation}   \triangle I=649.4 \cdot F^{0.7} \end{equation}

\section{SUMMARY}
\label{GLOB-method}

This paper suggests a new method for investigating the ionospheric
response to faint solar flares (of X-ray class C) when the
variation amplitude of the TEC response to individual LOS is
comparable to the level of background fluctuations. The dependence
of the TEC variation response amplitude on the flare location on
the Sun is investigated. In the case of solar flares whose class
is similar in X-ray emission, the dependence under study
resembles cos(CMD). The high sensitivity of our method permits us
to propose the problem of detecting, in the flare X-ray and EUV
ranges, emissions of non-solar origins which are the result of
supernova explosions. For powerful solar flares it is not
necessary to invoke a coherent summation, and the ionospheric
response can be investigated for each beam. This opens the way to
a detailed study of the SID dependence on a great variety of
parameters (latitude, longitude, solar zenith angle, spectral
characteristics of the emission flux, etc.). With current
increasing solar activity, such studies become highly
challenging. In addition to solving traditional problems of
estimating parameters of ionization  processes in the ionosphere
and problems of reconstructing emission  parameters, the data
obtained through the use of our method can be used to estimate the
spatial inhomogeneity of emission fluxes at scales of the Earth's
radius.
\section{Acknowledgements}

Authors are grateful to E.A. Kosogorov and O.S. Lesuta for
preparing the input data. Thanks are also due V.G. Mikhalkovsky
for his assistance in preparing the English version of the
manuscript. This work was done with support under RFBR grant of
leaping scientific schools of the Russian Federation No.
00-15-98509 and Russian Foundation for Basic Research (grants
99-05-64753,00-05-72026, 00-07-72026 and 00-02-16819a), GNTP
'Astronomy'.

\end{document}